# Time Warps in Photonic Time Stretch ADC and Their Mitigation

Shalabh Gupta and Bahram Jalali
Department of Electrical Engineering, University of California, Los Angeles
Los Angeles, CA 90095, USA
Email: {shalabh, jalali}@ee.ucla.edu

*Abstract*— We demonstrate a 10-GHz bandwidth digitizer with 7-effective bits of resolution and 52dB SFDR, using photonic time stretch technique. To the best of our knowledge, this is the highest resolution analog-to-digital converter (ADC) of the same bandwidth, with at least an order of magnitude higher SNR than previously achieved. This is made possible by correction of distortion due to time warps, and non-linearities due to wavelength dependent bias variation in the Mach-Zehnder modulator. We also demonstrate concatenation of 30 wavelength interleaved time segments with high fidelity on the path to achieving continuous time operation.

## I. INTRODUCTION

The need for high performance ADCs has steadily been growing as they form the unique interface between the analog and the digital world. With demand for increased bandwidths from the internet backbone, researchers are targeting 100-Gb/s and higher data rates per WDM channel to meet these requirements. To achieve this, multi-level optical modulation [1] is proposed, but it requires very high performance ADCs at the receiver end. In the wireless domain also, rapid advancement in capabilities of digital electronics due to CMOS scaling has pushed for the demand for high bandwidth and high resolution ADC front-ends, as in case of Software Defined Radios [2]. Also, there is a key requirement for such ADCs for defense applications like radars, and in research for better test and measurement systems. All these applications make the need for high performance and resolution ADCs insatiable.

Even with the advancement in CMOS electronics, ADCs have poor resolution at high RF frequencies. The jitter in the sampling clocks and comparator uncertainties act as a barrier to obtaining good SNR (signal to noise ratio) at multi-GHz RF frequencies [3]. By using photonic Time-Stretch pre-processing [4]-[5], this jitter barrier can be overcome and ultra-fast waveforms can be digitized [6]. However, the emphasis until recently had been on improving the sampling rates using time-stretch technique with single shot operations. In this paper, we demonstrate the use of this technique for capturing signals with high resolution. We also show two WDM (wavelength division multiplexed) channel operation to illustrate scalability to continuous time system.

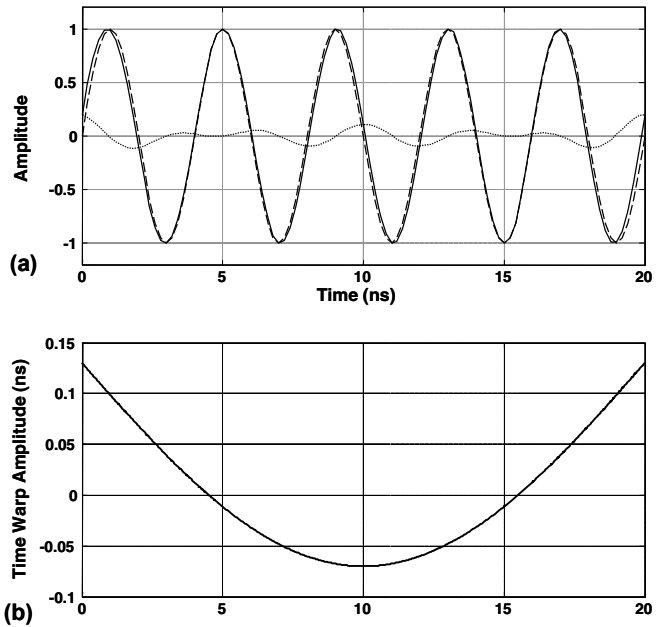

Figure 1. (a) Solid curve shows a signal with time warp; dashed curve is the ideal signal without any time-warp (a perfect sinusoid); the dotted curve shows the error (their difference). (b) The amplitude of the time warp

In the photonic time stretch ADC (TS-ADC), the RF signal to be digitized is modulated over a chirped optical pulse, and is subsequently time-stretched by passing it through a dispersive medium. The time stretched optical field is converted back to the electrical signal at a photodetector. This gives a slow replica of the original RF signal that can be captured by commercially available electronic digitizers, and hence, enabling capture of very high bandwidth electrical signals [4]-[5]. Additionally, stretching reduces the noise bandwidth of the signal and also relaxes jitter requirements on the electronic clocks, thereby enabling higher SNRs [7].

After quantization using ADC, signal processing is performed in digital domain to remove envelope modulation and correct for non-linear distortions due to Mach-Zehnder modulator transfer function and dispersion induced non-

This work was supported by DARPA under SPAWAR grant No. N66001-07-1-2007.).

linearity [8]-[9]. This processing corrects for most of the intermodulation and harmonic distortion in the signal.

However, the waveforms thus recovered, still have a slow varying RF phase distortion or "time warp" as shown in Fig. 1. This time warp can be caused by non-uniform time stretch whose origins are discussed in the next section. In this section, we also discuss how stitching of the waveforms from different wavelength channels is performed by removing skews between the channels with high precision, in the same way as removing the time warps.

As discussed in [10], the bias offset in a Mach-Zehnder modulator can drift significantly with the wavelength. If the bias offset is slow varying, it can be removed during arcsine linearization and high pass filtering [8]. However, the continuous time implementation of the TS-ADC is expected to have a large number of wavelength channels covering very wide optical bandwidth with the bias offsets varying at the repetition rate of the mode-locked laser. In Section 3, we discuss non-linear distortions due to these effects, and also show how they are removed. These sections are followed by experimental setup description, results and conclusions.

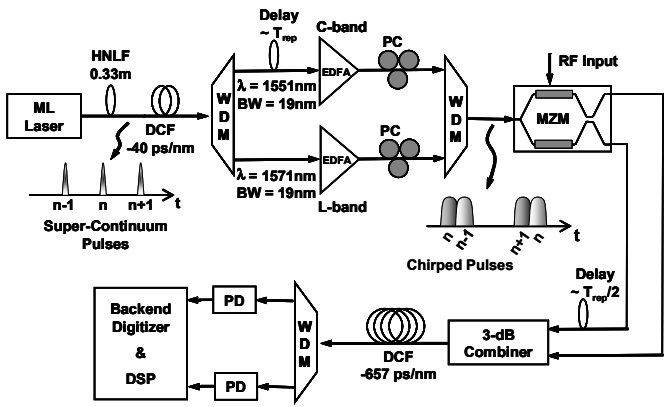

Figure 2.  Experimental Setup

## II. TIME WARP AND CHANNEL SKEWS

In a time stretch ADC, the stretch or the magnification factor is given by $M=(D_1+D_2)/D_1$, where $D_1$ and $D_2$ are the dispersion values of the first and second DCFs [7], as shown in Figure 2. In this system, it is assumed the non-linear dispersion coefficients in the two fibers (i.e. $\beta_3$, $\beta_4$ etc.) are either negligible or at least their mismatches in the two fibers are negligible. In such a scenario, the stretch factor of the system remains constant and independent of the optical wavelength [7]. However, this may not always be true, particularly in case of very wide optical bandwidths, and the stretch factor becomes wavelength dependent. Since there is a direct wavelength-to-time in the TS-ADC, this wavelength dependent stretch factor leads to the time warp.

In addition to non-linear dispersion, there are other non-idealities that can result in significant amount of time warps. The ultra-fast optical pulses originating from the mode locked laser initially have very high peak power. As they go through initial few hundred meters of the DCF, they may experience self-phase modulation (SPM) due to optical Kerr non-linearity in the fiber. After traversing some distance, the pulse widths become very broad and peak power drops and SPM is no longer significant. However, SPM broadening in the initial stage of the 1st DCF can significantly distort the time-wavelength mapping culminating in the time warps.

It is well known that the WDM filters have a small amount of wavelength dependent group delay variation [12], which can also become a significant source of time warps. In the TS-ADC, however, if these filters are placed after the Mach-Zehnder modulator, the effect of this variation is reduced by the stretch factor.

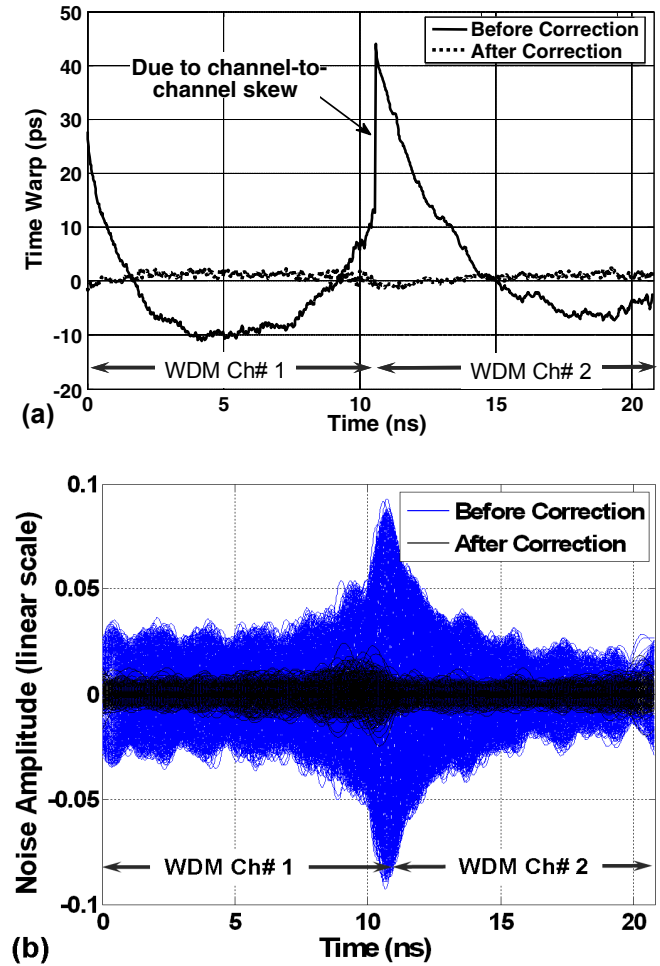

Figure 3.  (a) Time warp magnitudes before and after correction along the time axis of two combined pulses. (b) Overlaid noise profiles for multiple pulses before and after correction.

In addition to time warps, when two or more channel waveforms are combined together, there are small timing skews between them. Even though most of the skews can be removed in hardware by arranging clock and delays in a feedback system, some skews might still remain. To the first order, these skews effectively appear as time warps at the boundaries of the channels.

A global solution is proposed here to correct for all these effects. First the RF signals are obtained by envelope correction explained in [8], giving sinusoids with time warps. The waveforms from adjacent WDM channels are coarsely aligned in time and concatenated. Sine curve fitting is used in these waveforms, and average zero crossing between the obtained waveforms and ideal sine-fit curves for a large number of such waveforms is plotted along the time axis, as shown in Fig. 3(a). Before correction, roughly quadratic time warps are seen in the two channels. The sharp jump in the time warp in the middle indicates the timing skew between sinusoids from the adjacent WDM channels. Obtained data is used to correct time warp distortion by performing $2^{nd}$ order polynomial interpolation. Fig. 3(b) shows the noise profiles of the waveforms before and after time warp correction.

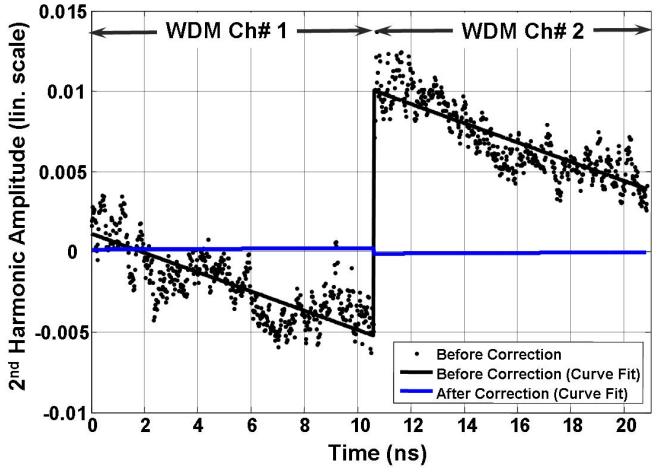

Figure 4. Second Harminc Amplitude as a function of time / wavelength

### III. MACH-ZEHNDER BIAS CORRECTION

It is well known that the Mach-Zehnder modulator bias offset drifts with wavelength because it is very difficult to perfectly match the two arms of the modulator [10]. Therefore, for wide optical bandwidths, this phenomenon becomes very problematic adding significant second order distortions, that vary along the time axis in the recovered waveforms. The problem can be solved to some extent by using differential and arcsine approach proposed in [8]. However, it is found that in this approach, estimating correct DC offsets becomes difficult because second order non-linearity itself generates extra DC terms. Also there are other common mode DC offsets that are not due to Mach-Zehnder bias. To overcome these issues, and to correct for non-linearity due to Mach-Zehnder bias offsets, same Sine-fitting technique was used as was done for time warp correction. However, in this case, the average amplitude of second harmonic is plotted along the time axis for the combined waveforms, as shown in Fig. 4. It was found that second harmonic amplitude $\delta(t)$ was drifting linearly with the wavelength, in accordance with our expectations. Once $\delta(t)$ is obtained, the linearized signal y(t) is obtained from the original signal x(t) as

$$y(t) = x(t) - \delta(t).(2x^2(t) - 1) \qquad (1)$$

An RF tone, $\cos(\omega t)$, with $2^{nd}$ harmonic amplitude profile given by $\delta(t)$, will take the form $x(t)=[\cos(\omega t) + \delta(t).\cos(2\omega t)]$. For this signal, $y(t)=[\cos(\omega t)+(\text{terms with coefficient } \delta^2(t))]$. Since $\delta(t)$ itself is small, terms with $\delta^2(t)$ become negligible and hence original signal $y(t) \sim \cos(\omega t)$ is obtained. After performing all these corrections, the "static" common mode noise caused by slow varying optical effects, which is common to all waveforms, is subtracted from them. Finally, arcsine linearization is performed [8], to suppress the third order non-linearities caused by Mach-Zehnder transfer function.

### IV. EXPERIMENTAL SETUP

The experimental setup consists of the time stretch system as shown in Fig. 2. Optical pulses generated by a passively mode locked laser (MLL) are sent through a highly non-linear fiber (HLNF) of length 0.33m. Self phase modulation in the fiber broadens the optical bandwidth of the pulses from about 17nm to more than 40nm, giving nice and flat super-continua in standard 1551nm and 1571nm telecom CWDM bands. These pulses are stretched using the first dispersion compensation fiber (DCF) with group velocity dispersion value of -40ps/nm, resulting in chirped optical pulses. Super-continua are carved out using the CWDM filters and the pulses are amplified in C-band and L-band band EDFAs (erbium doped fiber amplifiers) for the two channels, respectively.

In addition to amplification, the 1551nm channel pulses are shifted in time by an additional patch cord, so that the trailing edge of $n^{th}$ pulse in the 1571nm band slightly overlaps with the leading edge of $(n-1)^{th}$ pulse in the 1551nm band. The small overlap is provided to ensure that the RF signal has continuity between adjacent WDM channel pulses. The overlapping L-band and C-band optical pulses are derived from separate laser pulses to demonstrate that the time stretched signals obtained from adjacent laser pulses can be combined seamlessly, as this will be a requirement for full continuous time operation.

Separate polarization controllers are used for the two WDM channels after the EDFAs to independently correct for their polarizations, before the optical pulses are combined in the second CWDM filter, and fed to the intensity modulator. Push-pull (zero-chirp) Mach-Zehnder modulator with dual outputs is used to modulate the RF signal over the optical pulses, to facilitate use of differential operation, which helps removing dispersion induced non-linearity and improving SNR by 3dB [8]. For ease of implementation, the two Mach-Zehnder modulator outputs are delayed and skewed in time with respect to each other and time division multiplexed onto the $2^{nd}$ DCF. The $2^{nd}$ DCF has dispersion value of -657ps/nm, which results in a stretch ratio of about 17. The WDM channels at the DCF output are demultiplexed and detected using amplified photo-detectors, which receive about 0.5mW optical power. The RF signal so obtained is digitized by a commercially available oscilloscope (DPO71604), and further signal processing is performed on a computer to reconstruct the original RF signal.

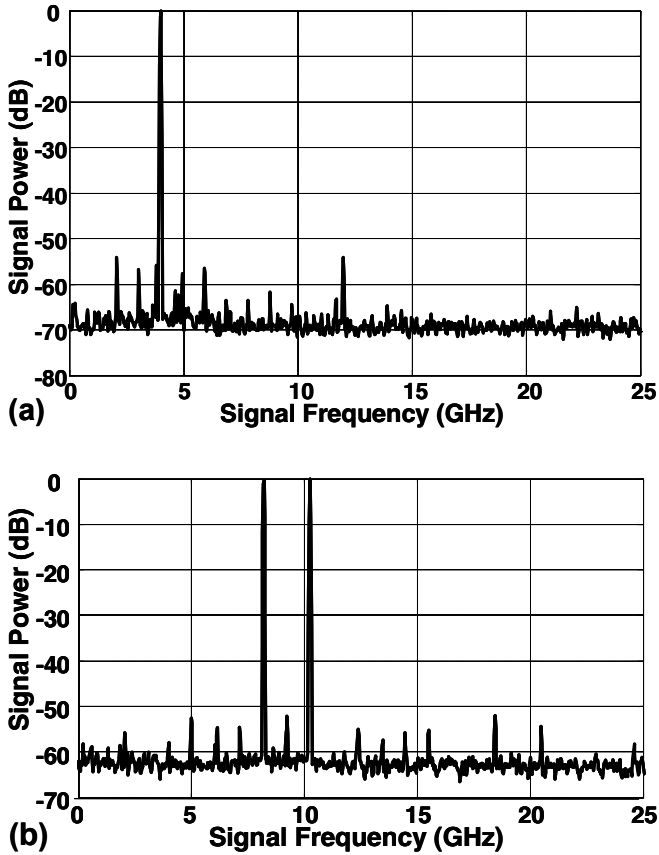

Figure 5. (a) Single tone test for a 4GHz tone (30 waveforms are stitched together coherently, to achieve high resolution bandwidth); ENOB = 7.22, SFDR=54dB. (b) 2-tone test for 8.2 and 10.25 GHz tones, ENOB=6.65, SFDR=52dB.

V. RESULTS

Tests were performed with single tone signals that were varied from 4GHz to 16 GHz with modulation depth of ~0.7, and also for two-tone signals with the modulation depth of ~0.5. It was found that the time warp distortion was almost constant over time, and channel-to-channel timing skews varied at a slow rate. However, Mach-Zehnder modulator bias offsets were varying much faster (in the order of a few seconds to a few minutes).

For any particular signal, first half of the captured waveforms (365 pulses in 10μsec) were used to obtain calibration data, which was applied for correcting distortions in the remaining 365 pulses. From this effective number of bits (ENOBs) of these corrected waveforms were obtained. On an average, single tone waveforms gave 7.05 to 7.25 ENOBs (obtained using sine fit method), with 10GHz noise bandwidth. For two tone tests, signal power had to be reduced by 3dB to avoid clipping, resulting in about 6.5 to 6.7 ENOBs. In other words, when referred to full scale, 7.0 to 7.2 ENOBs were obtained for 2-tone tests. To obtain SFDR, high resolution bandwidth was required, for which 30 segments were stitched synchronously with high fidelity. Different tests gave the worst case SFDR of 52dB. Fig. 5 shows the FFT plots of the obtained waveforms for two different cases. The obtained SNR was limited primarily by the resolution of the backend digitizer, which is about 7.2 ENOBs over 1GHz bandwidth.

VI. CONCLUSIONS

We have demonstrated the capture of 10GHz bandwidth RF signals with >7 effective bits of resolution and >52dB SFDR – a world record in ADC performance. This has been made possible by use of photonic time stretch technique and use of a new powerful algorithm to correct for distortions caused by time warps and Mach-Zehnder wavelength dependent bias variation. We have also demonstrated a path to continuous time operation by concatenation of 30 time segments while maintaining high signal fidelity.

ACKNOWLEDGMENT

The authors would like to thank George C. Valley, Gary Betts and Ozdal Boyraz for highly useful discussions, and Daniel Solli for providing highly non-linear optical fiber.